\documentclass[showpacs, twocolumn, prb, citeautoscript]{revtex4}
\usepackage{graphicx}
\usepackage{times}

\newcommand{\be}{\begin{equation}}
\newcommand{\ee}{\end{equation}}
\newcommand{\ba}{\begin{eqnarray}}
\newcommand{\ea}{\end{eqnarray}}
\newcommand{\Delc}{\Delta_{\rm c}}

\begin{document}

\title{Quantum Monte Carlo Study of Disordered Fermions}

\author{Ji-Woo Lee, Shailesh Chandrasekharan, and Harold U. Baranger}
\affiliation{Department of Physics, Duke University, Durham, 
North Carolina, 27708-0305}

\date{16 May 2005; \textbf{Phys. Rev. B 72, 024525 (2005)}; DOI: 10.1103/PhysRevB.72.024525}

\begin{abstract}

We study a strongly correlated fermionic model with attractive interactions
in the presence of disorder in two spatial dimensions. Our model has been 
designed so that it can be solved using the recently discovered meron-cluster 
approach. Although the model is unconventional it has the same symmetries as 
the Hubbard model. Since the naive algorithm is inefficient, we develop a 
new algorithm by combining the meron-cluster technique with the directed-loop 
update. This combination allows us to compute the pair susceptibility and 
the winding number susceptibility accurately. We find that the s-wave 
superconductivity, present in the clean model, does not disappear until 
the disorder reaches a temperature dependent critical strength. The critical
behavior as a function of disorder close to the phase transition belongs to 
the Berezinky-Kosterlitz-Thouless universality class as expected. The 
fermionic degrees of freedom, although present, do not appear to play an 
important role near the phase transition.

\end{abstract}

\maketitle

\section{Introduction}

A variety of systems show quantum coherence over large distances at 
low temperatures. Superfluidity and superconductivity are two striking
physical phenomena showing such behavior, which have been extensively 
studied over the years. However, when correlations between the microscopic 
degrees of freedom become strong it is difficult to study these phenomena
theoretically from first principles. The calculations must take into
account strong fluctuations over many length scales which is only possible 
numerically. When the microscopic degrees of freedom involve bosonic
variables one can usually devise efficient quantum Monte Carlo methods to 
solve the problem\cite{Cep95}. On the other hand, it is still difficult to 
study a variety of models from first principles when the 
microscopic theory is fermionic. For example, the critical temperature 
below which superconductivity is seen in the well-known attractive Hubbard 
model was only determined recently\cite{Pai04}, using the determinantal Monte
Carlo method\cite{Whi89,Jar96} and on lattices only as large as $18\times 18$. 
The main approaches to dealing with fermionic systems can be viewed as arguments that universality allows one to replace the microscopic theory with an effective low energy theory.  The resulting effective theory is usually either a Fermi-liquid theory, a BCS-type mean-field theory, or some bosonic theory\cite{PhillipsBook}. 
A key element in furthering microscopic understanding, then, is to validate the universality arguments and determine the low-energy effective theory; in practice, this has proved very difficult for systems with strong correlations.

Real systems usually contain impurities. Thus, in addition to understanding
superconductivity in clean systems, the effects of impurities in the form of
disorder need to be incorporated in the studies. In certain systems like 
two-dimensional superconducting films and Josephson-junction arrays, it has 
been discovered that superconductivity can be destroyed by tuning parameters 
such as the film thickness \cite{Goldman1,Goldman2}. Since these tuning 
parameters change the effective strength of the disorder, it is believed that 
the superconductor-to-insulator (SI) phase transitions in these systems
can be understood as being driven by disorder. Among the models used to 
explain the experiments, the attractive Hubbard model with disordered
chemical potential is one well-known starting point \cite{Litak92,Scalettar}. 

The relevance of disorder for superconductivity was first addressed by 
Anderson\cite{Anderson}, where he argued that superconductivity is
insensitive to perturbations that do not destroy time reversal 
invariance. Using a BCS type trial wave function Ma and Lee\cite{Ma} 
showed that superconductivity can persist even below the mobility edge.
Clearly, these studies suggest that an SI transition is an effect of
strong disorder which makes it a difficult subject for analytic work.
Fisher {\it et al.}\cite{Fisher} have argued that the effective theory describing the transition is bosonic, and then developed a deeper understanding of the purely bosonic superfluid-insulator transition using renormalization group arguments along with scaling.
A variety of quantum Monte Carlo work has been done over the years on these purely bosonic microscopic theories\cite{Sca91,Kra91,Wal94,Lee01,Ber02,Pro04}. 
If fermions do not play an important role 
near the transition, it is likely that these studies will also be useful 
in understanding the universality of the fermionic SI transition. Recently determinantal quantum Monte Carlo studies of the attractive (fermionic) Hubbard 
model with disorder have been performed\cite{Scalettar}, 
which show that it is indeed possible to drive an SI transition by 
increasing the disorder and, as expected, the critical disorder is large.  
However, the system sizes explored were quite small, $8\times 8$. Other
studies of disorder effects also involved only very small systems
\cite{Sri02,Den03}.

Motivated by the physics of the SI transition, in this article we study 
the effects of disorder in a strongly correlated fermionic model. Our model
is unconventional and has been built so that it can be studied using the 
recently discovered meron cluster algorithms\cite{meronPRL99,Shailesh}. These novel 
algorithms are so efficient that lattices as large as $L \!=\! 128$ were studied
recently, and it was shown with great precision both that the low temperature 
phase of the clean model is indeed superconducting and that the finite 
temperature phase transition belongs to the Berezinski-Kosterlitz-Thouless 
(BKT) universality class\cite{Shailesh02,Osb02}. Here we use the same
model to explore the effects of disorder on superconductivity
and, in particular, focus on the role of fermions. Unfortunately, the naive extension 
of the earlier algorithm becomes inefficient in the presence of disorder; 
hence, we also develop a new algorithm by combining the 
meron-cluster formulation with the directed loop algorithm\cite{Sandvik}. 
This new algorithm allows us to measure the relevant observables very 
accurately.

Our paper is organized as follows: In Section II, we introduce
our model and define the observables that we use later.
In Section III, we rewrite the model in a cluster representation.
Section IV explains the new directed-loop algorithm we have 
developed. Section V contains our results, and Section VI contains our
conclusions and directions for the future.

\section{THE MODEL}

The model we consider in this article was motivated by the ability to solve 
the fermion sign problem using the the meron-cluster algorithm\cite{Shailesh}. 
The Hamiltonian of the model can be written as
\be
H = \sum_{\langle ij \rangle} H^{(2)}_{ij} + \sum_i H^{(1)}_i,
\ee
where $H^{(2)}_{ij}$ consists of all the nearest neighbor interactions
between sites $i$ and $j$ on an $L\times L$ square lattice and 
$H^{(1)}_{i}$ includes interactions on the site $i$. The term $H^{(1)}_i$ 
is given by
\ba
H^{(1)}_i &=& 
(U+J_3 - 1)(n_{i\uparrow}-\frac{1}{2})(n_{i\downarrow}-\frac{1}{2}) 
\nonumber \\
&& - \frac{\mu_i}{2} (n_{i\uparrow} + n_{i\downarrow} -1 )
\ea
where $U$ represents the Hubbard interaction between spin-up and spin-down 
electrons and $\mu_i$ is the local chemical potential, through which disorder 
is introduced in the model. The term $H^{(2)}_{ij}$ is unconventional and is 
given by
\ba
H^{(2)}_{ij} &=&  \frac{1}{4}
(c^{\dagger}_{i\sigma}c_{j\sigma} + c^{\dagger}_{j\sigma} c_{i\sigma} )
		\{(1+J_3)( n_{ij}^2 - 4 n_{ij} + 3 ) \nonumber \\
		& & - (1-J_3) ( n_{ij} - 2 )\}  \nonumber\\
		&+& \{ {\bf S}_i \cdot {\bf S}_j +  {\bf J}_i \cdot {\bf J}_j -
			(1-J_3) J^z_i J^z_j \} \\
&- & \frac{1+J_3}{4}(n_{i \uparrow} - \frac{1}{2} ) (n_{i \downarrow} -
		\frac{1}{2} ) 
	(n_{j \uparrow} - \frac{1}{2} ) (n_{j \downarrow} - \frac{1}{2} )
	\nonumber
\ea
Here $c^\dagger_{i\sigma}$ and $c_{i\sigma}$ are the usual creation 
and destruction operators of spin $\sigma$ at site $i$, $n= c^\dagger c$, and
$n_{ij}=\sum_\sigma n_{i\sigma} + n_{j\sigma}$. $\mathbf{S}_i$ is the
spin operator on site $i$ defined by  
\be
\mathbf{S}_{i} = \frac{1}{2} \sum_{s, s'} c^\dagger_{is} \vec{\sigma}_{ss'} c_{is'}
\ee 
and $\mathbf{J}_i$ is the pseudo-spin operator defined by 
\ba
J^+_i &=& (-1)^{i_x+i_y} c^\dagger_{i\uparrow}c^\dagger_{i\downarrow}
\nonumber \\
J^-_i &=& (-1)^{i_x+j_y} c_{i\downarrow} c_{i\uparrow},
 \\
J^z_i  &=& \frac{1}{2} (n_{i\uparrow} + n_{i\downarrow} - 1). \nonumber
\ea
$J^+$ and $J^-$ are related to pair creation and annihilation operators.
In our notation $i_{x(y)}$ refers to the $x(y)$ component of the site $i$.

Although the Hamiltonian we study is unconventional, it has all the relevant
symmetries of the Hubbard model when $J_3 \!=\! 1$. In particular when $\mu \!=\! 0$ the 
Hamiltonian is invariant under the $SU(2)$ spin and $SU(2)$ pseudo-spin 
transformations. When $\mu\neq 0$, the pseudo-spin symmetry is broken to
the $U(1)$ fermion number symmetry. One can introduce repulsion or attraction
by making $U$ sufficiently positive or negative respectively. The important 
difference with the Hubbard model is that when $U \!=\! 0$ the model is still 
strongly interacting and by setting $J_3 \neq 1$ we can break the pseudo-spin 
symmetry. Further, the model simplifies in the $U\rightarrow -\infty$ limit; 
in this limit the model can be mapped to the simple Hamiltonian
\be
H= \sum_{\langle ij \rangle} 
{\bf J}_i \cdot {\bf J}_j + (J_3-1)J^z_iJ^z_i.
\ee
Clearly, when $J_3=1$, one obtains the antiferromagnetic Heisenberg model 
involving pseudo-spins, while $J_3 \!=\! 0$ leads to the XY model. When $J_3 \!=\! 1$,
$U\rightarrow\infty$ and $\mu_i \!=\! 0$ one gets the antiferromagnetic spin model
\be
H= \sum_{\langle ij \rangle} 
{\bf S}_i \cdot {\bf S}_j.
\ee
An interesting aspect of this model is that the fermion sign problem 
can be solved using the meron cluster approach for $0\leq J_3 \leq 1$
when $U < 0$ at any value of $\mu_i$. Also when $J_3 \!=\! 1$ the sign problem
can be solved when $U > 0$ when $\mu < U/2$. Thus, we think the model 
offers a rich phase diagram and deserves to be investigated. In this paper 
we will consider $-\infty < U \leq 0$ and investigate the physics when 
$J_3 \!=\! 0$ and $J_3 \!=\! 1$. We introduce disorder through $\mu_i  \!=\!  \mu + \delta\mu_i$
where $\delta\mu_i$ is a random number distributed uniformly from $-\Delta$ 
to $\Delta$.

In order to probe superconductivity in this system we will focus on two 
observables. The first is the pair susceptibility defined as
\be
\chi_{\rm p} = \frac{T}{V} \sum_{i,j} \int_0^{1/T} d\tau \int_0^{1/T} d\tau'
[P^+_{i\tau, j\tau'} + 
P^-_{i\tau, j\tau'}]
\ee
where $V$ is the spatial volume, $T$ is the temperature, and 
\be
P^+_{i\tau, j\tau'} = \frac{1}{4Z}
\mathrm{Tr} \Big\{
\mathrm{e}^{-(\beta-\tau) H} 
c^\dagger_{i\uparrow}c^\dagger_{i\downarrow}
\mathrm{e}^{-(\tau - \tau') H}
c_{j\downarrow}c_{j\uparrow} \mathrm{e}^{-\tau' H}\Big\}
\ee
is the pair correlation with 
$P^-_{i\tau,j\tau'} = P^+_{j(\beta-\tau'),i(\beta-\tau)}$. 
The second is the winding number susceptibility defined as  
\be
\chi_{\rm w} = \frac{\pi}{4 V} \sum_{i,j} 
\int_0^{1/T} d\tau \int_0^{1/T} d\tau'
[C^{(x)}_{i\tau,j\tau'} + C^{(y)}_{i\tau,j\tau'}]
\ee
where
\be
C^{(\mu)}_{i\tau,j\tau'}(\tau) = \frac{1}{Z}\mathrm{Tr} 
\Big\{\mathrm{e}^{-(\beta-\tau) H} 
J_\mu(i)\mathrm{e}^{-(\tau-\tau') H} 
J_\mu(j)\mathrm{e}^{-\tau'H} \Big\}
\ee
is the current-current correlation function where $J_\mu(i)$ is the conserved 
fermion current,
\be
J_\mu(i) = \frac{1}{2}\sum_{s=\uparrow,\downarrow} 
\Big[c^\dagger_{is}c_{(i+\hat{\mu})s} - c^\dagger_{(i+\hat{\mu})s}c_{is}\Big],
\ee
 at the site $i$ in the $\hat{\mu} = \hat{x},\hat{y}$ direction. 

In order to estimate the importance of fermions we will also look at the 
density of singly occupied sites defined as
\be
n_s = \frac{1}{V Z} \sum_i \mathrm{Tr}
\Big[ \exp(-\beta H) \{
n_{i\uparrow} + n_{i\downarrow} - 2 n_{i\uparrow} n_{i\downarrow}\}\Big]
\label{ns}
\ee
and at the total density of electrons defined by
\be
n = \frac{1}{V Z} \sum_i \mathrm{Tr}
\Big[ \exp(-\beta H) \{n_{i\uparrow} + n_{i\downarrow}\}\Big].
\label{n}
\ee
A comparison of $n_s$ and $n$ will tell us how many sites have formed 
local pairs.

\section{CLUSTER REPRESENTATION}

\begin{figure}[b]
\includegraphics[width=6cm]{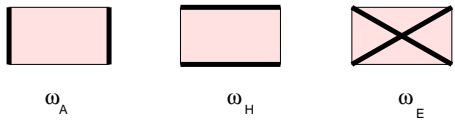}
\caption{\label{weights} The three bond configurations on space-time
plaquettes and their weights. The values of the weights are given in
Eq.~(\ref{pwt}).}
\end{figure}

\begin{figure}[t]
\includegraphics[width=6cm]{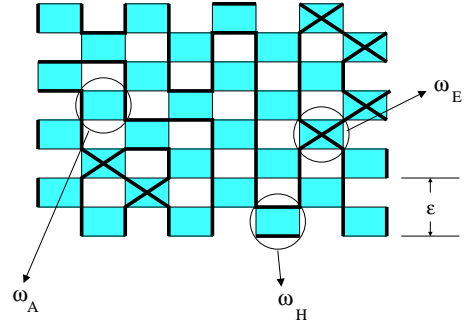}
\caption{\label{cconf} An example of a bond configuration in one space and
one time dimension. The configuration contains five clusters.}
\end{figure}

It is possible to rewrite the partition function of our model in discrete 
time in terms of a statistical mechanics of closed loops on a space-time 
lattice\cite{Shailesh}. We first divide 
$\beta$, the length in the Euclidean time direction, into $M$ equal steps 
such that $\epsilon = \beta/M$. Interactions between nearest neighbor 
sites are introduced in a checker-board type manner, so that on each 
time slice every site interacts with a unique neighbor. This then introduces 
$4$ extra time slices for every $\epsilon$ time step. In the cluster
representation the nearest neighbor interactions occur in the form of
three types of bond configurations on space-time plaquettes as shown 
in Fig.~\ref{weights}. Their weights are given by,
\ba
\omega_A 
&=& e^{\epsilon J_3/4} ( e^{-\epsilon J_3 /2 }+ e^{-\epsilon/2} )/2 
\nonumber \\
\omega_H 
&=&  e^{\epsilon J_3/4} ( - e^{-\epsilon J_3/2}+ e^{\epsilon/2} )/2 
\label{pwt}
 \\
\omega_E 
&=&  e^{\epsilon J_3/4} ( e^{-\epsilon J_3/2} - e^{-\epsilon/2} )/2 \nonumber
\ea
Given a configuration $\{C\}$ of bonds, one can connect them together to 
form many closed loops; we denote them by ${C_\alpha},\alpha=1,2,...,N_c$. 
The partition function can then be written as
\be
Z = \sum_{C} \Big\{\prod_\alpha \Omega(C_\alpha) 
\Big(\prod_{\cal P} \omega_{\cal P}\Big)\Big\}
\ee
where $\omega_{\cal P}$ is the weight of the bond associated with the
plaquette ${\cal P}$ and takes one of the values given in Eq.~(\ref{pwt}).
\be
\Omega(C_\alpha) = 2 \cosh ( \frac{\epsilon}{8} \mu_{C_\alpha} ) 
+ \sigma(C_\alpha) 2 e^{ \frac{\epsilon}{2} \frac{U}{4} S_{C_\alpha}}
\label{clwt}
\ee
is the weight corresponding to each loop $C_\alpha$ that arises due to the 
fermionic degrees of freedom associated with the loop. Here 
\be
S_{C_\alpha} \equiv \sum_{(i\tau) \in C_\alpha} 1
\ee
where the sum is over all all space-time points that belong to the cluster
$C_\alpha$. Thus $S_{C_\alpha}$ is just the size of the cluster.
On the other hand
\be
\mu_{C_\alpha} \equiv \sum_{(i\tau) \in C_\alpha} \mu_i w_{i\tau},
\ee
where $w_{i\tau}$ is $+1$ when the cluster is going forward in time and 
$-1$ when it is going backward in time at the site $i\tau$. If the
cluster moves horizontally, our convention is that the temporal direction is 
reversed. In order to determine $w_{i\tau}$ one can start from any 
point and traverse the cluster in either direction. Note that
\be
W_t(C_\alpha) \equiv \sum_{(i\tau) \in C_\alpha} w_{i\tau}.
\ee
is the temporal winding of the cluster $C_\alpha$. Finally, the factor 
$\sigma(C_\alpha)$ in Eq.~(\ref{clwt}) is a sign factor associated with the 
cluster topology, that arises due to the fermion permutation signs.
\cite{Shailesh} Following Ref.~{~}\onlinecite{Shailesh} 
we call the cluster 
a meron if $\sigma(C_\alpha) = -1$. If $N_h(C_\alpha)$ is the number of
horizontal hops in the cluster $C_\alpha$, then the cluster is a meron 
if and only if $N_h(C_\alpha)/2 + W_t(C_\alpha)$ is even. An example of a 
bond configuration in two dimensions is shown in Fig.~\ref{cconf}.

In the cluster representation it is easy to show that
\ba
\lefteqn{[P^+_{i\tau,j\tau'} + P^-_{i\tau,j\tau'}] = }
\nonumber \\
&& \Big\langle 
\sum_\alpha 
\frac{2 \cosh \Big(\frac{\epsilon}{8} \mu^{(i\tau,j\tau')}_{C_\alpha}\Big)
}{\Omega(C_\alpha)} 
\delta_{(i\tau;j\tau')\in C_\alpha} \Big\rangle
\label{pcorr}
\ea
where $\delta_{(i\tau;j\tau')\in C_\alpha}$ imposes the restriction that 
both the space-time sites $(i\tau)$ and $(j\tau')$ belong to the 
cluster $C_\alpha$ and
\be
\mu^{(i\tau,j\tau')}_{C_\alpha} = \sum_{(k\tau'') \in C_\alpha} 
s^{(i\tau,j\tau')}_{k\tau''} \mu_i w_{k\tau''}
\ee
where $s^{i\tau,j\tau'}_{k\tau''}$ is $+1$ while going from $(i\tau)$ to 
$(j\tau')$ and $-1$ while continuing from $(j\tau')$ to $(i\tau)$. 
The winding number susceptibility is given by
\ba
\chi_{\rm w} &=& \frac{\pi}{4 V} \sum_\mu \Bigg\langle \Bigg\{ 
\sum_{\alpha} 
\frac{W^2_{\mu\alpha}\cosh ( \frac{\epsilon}{8} \mu_{C_\alpha})}
{2 \Omega(C_\alpha)}
\nonumber \\
&+& \sum_{\alpha \neq \alpha'} 
\frac{W_{\mu\alpha}W_{\mu\alpha'}
\sinh ( \frac{\epsilon}{8} \mu_{C_{\alpha}})
\sinh ( \frac{\epsilon}{8} \mu_{C_{\alpha'}})}
{\Omega(C_{\alpha})\Omega(C_{\alpha'})}
\Bigg\}\Bigg\rangle
\nonumber \\
\label{chiw}
\ea
where $W_{\mu\alpha}$ refers to the spatial winding of the loop clusters
along the spatial direction $\mu$. The density of single occupation turns 
out to be
\be
n_s = \frac{1}{4\ M\ V} \Big\langle \sum_{\alpha} 
\frac{\sigma(C_\alpha) S_{C_\alpha} 
2 e^{\frac{\epsilon U}{4}S_{C_\alpha}}}{\Omega(C_\alpha)} \Big\rangle
\label{cns}
\ee
and is a measure of the number of unpaired fermions.

\section{DIRECTED-LOOP ALGORITHM}

\begin{figure}[b]
\includegraphics[width=6cm]{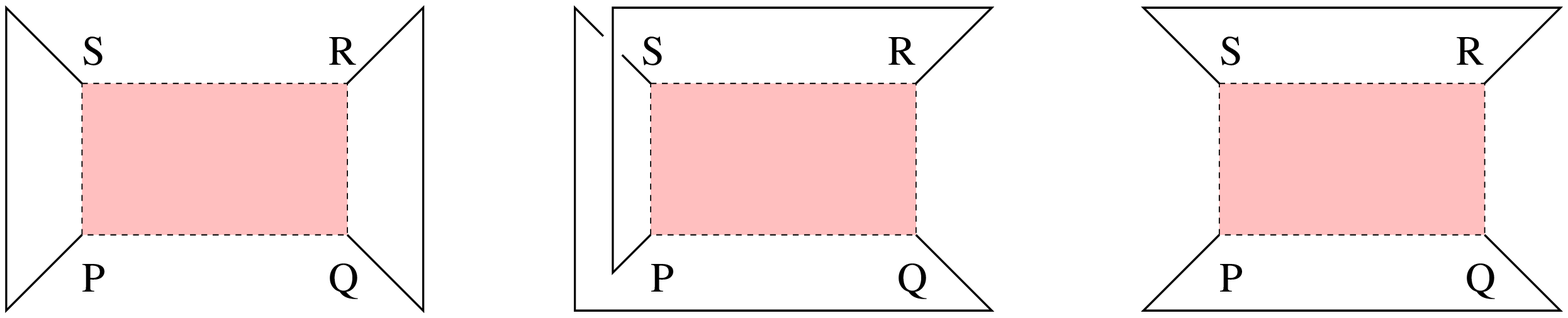}
\caption{\label{conn.fig} The three possible types of connectivity of the 
sites of a plaquette ($P,Q,R,S$) due to bonds on other plaquettes. }
\end{figure}

A simple Monte Carlo algorithm for the current problem involves visiting 
every interaction plaquette and updating the bond configuration on it by
replacing it with one of the three choices shown in Fig.~\ref{weights}. 
Since the Boltzmann weight also depends on the structure of the loops 
formed by these bonds, the decision involves figuring out the connectivity 
of the sites of the plaquette (referred to as $P,Q,R,S$) due to the bonds 
on other plaquettes. This connectivity can be one of three types as shown in 
Fig.~\ref{conn.fig}. Thus, choosing a new interaction involves finding
weights of nine configurations (three bond configurations for three types
of connectivity) and the new bond configuration can be found by using a heat 
bath or a Metropolis step.

We have found that in the presence of a chemical potential $\mu$, this 
simple algorithm is inefficient. This behavior can be understood by noting
that a chemical potential is similar to a magnetic field in a
a quantum spin model and in the context of quantum spin models
there is evidence that this type of naive algorithm becomes inefficient
in the presence of magnetic fields \cite{Cha99,Tro03}. Today it is well
known that quantum spin systems in the presence of a magnetic field 
can be solved efficiently using the directed-loop algorithm\cite{Sandvik}. 
However, until 
now this algorithm has been formulated only in the spin representation 
and not in the cluster representation. Unfortunately, the sign problem 
in the fermionic model can only be solved in the cluster representation.
In this article we show how one can extend the directed-loop algorithm 
to the cluster representation which then leads to an efficient algorithm 
for the fermionic model even in the presence of a chemical potential.

The basic idea behind the directed loop algorithm is to extend the 
configuration space so that configurations that contribute to certain 
two point correlation functions (denoted $\{C^{(2)}_{i\tau,j\tau'}\}$) 
are sampled along with the configurations that contribute to the partition 
function (denoted $\{C\}$). The configurations $C^{(2)}_{i\tau,j\tau'}$ 
have two reference space-time points, $i\tau$ and $j\tau'$; 
during the directed-loop update one of these points, say $i\tau$, is held 
fixed while the other point $j\tau'$ is moved around. The directed loop 
update begins with a configuration in the set $\{C\}$ and chooses a site 
$i\tau$ at random and probabilistically creates a configuration in the 
set $\{C^{(2)}_{i\tau,j\tau'}\}$, with $i\tau=j\tau'$. The probability of 
creating this configuration must satisfy detailed balance in order to
produce configurations $\{C\}$ and $\{C^{(2)}_{i\tau,j\tau'}\}$ with the
correct Boltzmann weight. Once a configuration in the set 
$\{C^{(2)}_{i\tau,j\tau'}\}$ is created, the 
point $j\tau'$ is moved around while satisfying detailed balance and thus 
sampling other configurations in the set $C^{(2)}_{i\tau,j\tau'}$ with
the correct Boltzmann weight. Finally,
when the two points meet again, i.e.., when $j\tau'=i\tau$, the two 
points may be removed to obtain a configuration in the set $\{C\}$ in 
accord with detailed balance. Thus, since at every step detailed balance
is satisfied, it is easy to show that the directed-loop update, which 
starts from a configuration in $\{C\}$ and ends on another configuration 
in $\{C\}$, satisfies detailed balance. During the loop update,
all the sites encountered contribute to the two point correlation function.

In the current work we have used the pair correlations to develop the 
directed loop algorithm. Thus, the configurations in 
$\{C^{(2)}_{i\tau,j\tau'}\}$ are the ones that contribute to 
the pair correlations in Eq.~(\ref{pcorr}). Thus, the weight of such a
configuration is taken to be
\be
2 \cosh \Big(\frac{\epsilon}{8} \mu^{(i\tau,j\tau')}_{C_{\alpha_0}}\Big) 
\Big\{\prod_{\alpha \neq \alpha_0} \Omega(C_\alpha) 
\prod_{\cal P} \omega_{\cal P}\Big\}
\ee
where the sites $i\tau$ and $j\tau'$ are forced to remain on the same
cluster $\alpha_0$. The weight of a configuration in the set $\{C\}$ is
\be
\Big\{\prod_\alpha \Omega(C_\alpha) \prod_{\cal P} \omega_{\cal P} \Big\}.
\ee
The essential steps of the directed loop algorithm are as follows:
\begin{itemize}
\item[(i)] Start with the initial configuration which belongs to the set 
$\{C\}$.
\item[(ii)] Select a space-time site $i\tau$ at random and propose
to create a configuration $C^{(2)}_{i\tau,j\tau'}$ assuming $j\tau'=i\tau$. 
Accept the proposal with probability 
\be
\mathrm{Min}\Bigg\{
\frac{2\cosh \Big(\frac{\epsilon}{8} \mu^{(i\tau,j\tau')}_{C_{\alpha_0}}\Big)}
{\Omega(C_{\alpha_0)}}, 1\Bigg\}
\ee
where $C_{\alpha_0}$ is the cluster which contains the site $i\tau$.
\item[(iii)] If the proposal is not accepted then the update is complete.
Otherwise we go on.
\item[(iv)] The site $j\tau'$ is moved to the next site by picking the 
plaquette it is connected to which is not the one just visited.
Since each site is connected to two plaquettes the plaquette is unique
unless one is at the beginning. In that case one chooses one of the
two randomly.
\item[(v)] Each plaquette update involves choosing one of ten possible 
configurations. These configurations depend on the connectivity of
the plaquette and an example is given in Fig.~\ref{update1}.
\begin{figure}[t]
\includegraphics[width=6cm]{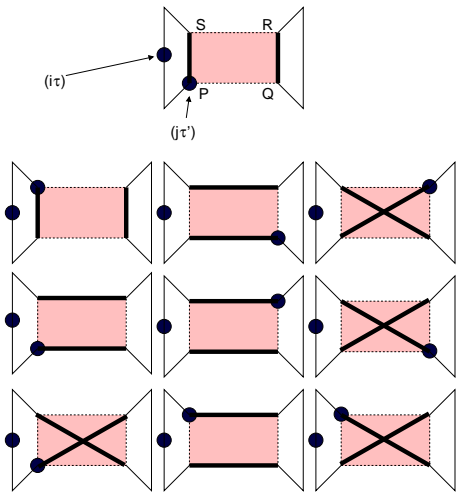}
\caption{\label{update1} Ten possible configurations for one of the
three types of connectivity shown in Fig.~\ref{conn.fig}. During
the directed loop update, depending on the connectivity of the
sites of the plaquette, a heat bath is used to pick one of ten 
possible configurations in order to move the site $(j\tau')$.}
\end{figure}
One of these ten is chosen using a heat bath.
\item[(vi)] Steps (iv)-(v) are repeated until $j\tau'$ reaches
$i\tau$, at which stage the transition to the $\{C\}$ sector is
made with probability
\be
\mathrm{Min}\Bigg\{
\frac{\Omega(C_{\alpha_0)}}
{2\cosh \Big(\frac{\epsilon}{8} \mu^{(i\tau,j\tau')}_{C_{\alpha_0}}\Big)}
,1\Bigg\}
\ee
If the transition is made then the directed-loop algorithm ends, otherwise
one goes back to step (iv) assuming one is in at the beginning of the
loop.
\end{itemize}
In the above algorithm, the pair susceptibility can be computed using
the formula
\be
\chi_{\rm p} = \frac{\epsilon}{16} \langle S \rangle
\ee
where $S$ is the number of sites visited during the directed loop update.
Other observables such as $\chi_{\rm w}$ and $n_s$ can be computed using
the formula of Eq.~(\ref{chiw}) and (\ref{cns}).

We have tested the efficiency of the directed loop algorithm by comparing the results with spin model results in the limit $U=-\infty$ which can be obtained using the usual directed loop algorithm. Since each update of the plaquette in the fermion algorithm requires knowing the connectivity of the loop, the algorithm is indeed much slower than the directed loop algorithm of a spin model which does not require this step. Unfortunately, this is a price one has to pay for being able to compute quantities in a fermionic theory. In Ref.~{~}\onlinecite{Shailesh02} a trick was used to reduce the time to determine the connectivity of the clusters. The trick was to use a ``tree'' structure to store the information about the cluster connectivity which allowed one to obtain the relevant information in a time that grew like the logarithm of the cluster size. This was not implemented in the current work but could be implemented if necessary.

\section{RESULTS}

In this Section we discuss the results obtained from extensive simulations for lattice sizes up to $L=32$. In our work we have fixed $\epsilon=0.25$ in order to avoid changes in the time discretization errors.  We have found that this value of $\epsilon$ is reasonably small and the results at smaller values appear to join within our error bars.  Further, since our desire is to understand universal physics of disorder, we believe that fixing $\epsilon$ should not be a major concern since it only changes the transfer matrix by a small amount. For a given disorder realization we typically discard the first $1000$ directed loop updates for equilibration and then average over $20000$ directed loop updates in blocks of $1000$ to generate each of our statistics. All quantities plotted at a given value of $\Delta$ have been averaged over $20$ disorder realizations.

\subsection{Superconductivity with Disorder}

It is known from earlier studies\cite{Shailesh02} that our model has a
low temperature superconducting phase when $U=0$, $J_3=0$ and 
$\mu=\Delta=0$. The superconductivity disappears at a finite temperature 
and the phase transition belongs to the expected BKT universality class. 
Here we study the effects of disorder on this 
system by keeping $\mu=0$ but $\Delta \neq 0$. In the clean model the BKT 
predictions for the leading finite size scaling form of the pair 
susceptibility and the winding number susceptibility are known:
\be
\chi_{\rm p}(L) = \left\{\begin{array}{cc}
 A L^{2-\eta} & T \leq T_c \cr
 A & T > T_c \end{array}\right.
\label{fitcp}
\ee
and
\be
\chi_{\rm w}(L) = \left\{\begin{array}{cc}
B \Big[ 1+ \frac{1}{2 \log \{ L/L_0 \}} \Big] & T \sim T_c \cr
B \exp(-L/L_0) & T \gg T_c \end{array}\right.
\label{fitcw}
\ee
where $A,B$ and $L_0$ are constants which depend on the temperature. We further expect $0\leq \eta \leq 0.25$, and $\eta=0.25$ with $B=2$ at the phase transition\cite{Shailesh02,Har97}. Here, instead of varying the temperature we fix the temperature at $T=0.25$ and study the effects of disorder by increasing its strength through the parameter $\Delta$. If the BKT universality holds we then expect the same finite size behavior to be true where the constants $A,B$ and $L_0$ now depend on $\Delta$.

\begin{figure}[t]
\vbox{
\includegraphics[width=7.7cm]{fig5-1.eps}
\includegraphics[width=7.7cm]{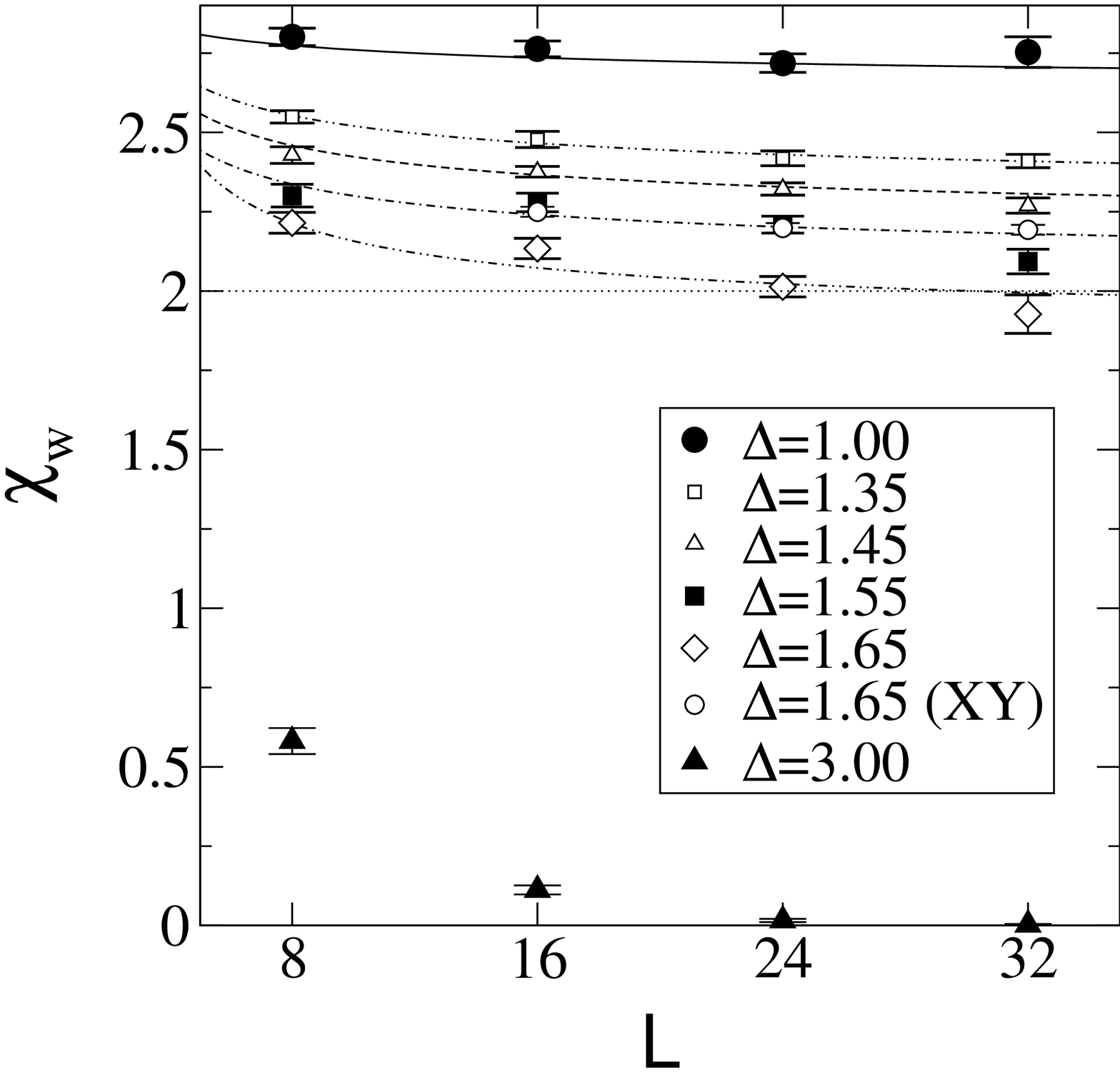}
}
\caption{ Pair susceptibility divided by $L^{1.75}$ ($\chi_{\rm p}/L^{1.75}$, top, logarithmic scales) and winding number susceptibility ($\chi_{\rm w}$, bottom, linear scales) as a function of lattice size for different magnitudes of disorder $\Delta$ at $J_3 \!=\! 0$, $U \!=\! 0$, $\mu \!=\! 0$, and $T \!=\! 0.25$. The unfilled diamonds give the value of $\chi_{\rm w}$ at $\Delta \!=\! 1.65$ in the $XY$ limit obtained when $U \!=\! -\infty$. The dashed horizontal lines are inserted as guides to the eye. The BKT transition appears to be near $\Delta \!\sim\! 1.5$.  \label{fig5}} \end{figure}

\begin{table}[t]
\vspace{0.05cm}
\caption{Fitting results for the pair susceptibility ($\chi_{\rm p}$) and winding number susceptibility ($\chi_{\rm w}$) 
for $J_3 \!=\! 0$, $U \!=\! 0$, and $\mu \!=\! 0$ (Fig.~\ref{fig5}).  
The fitting formulas are Eqs.~(\ref{fitcp})-(\ref{fitcw});
the $\chi^2$ given is per degree of freedom from the fit.}
\begin{center}
\begin{tabular}{l ||c c c| c c c       }
\hline
          &       & $\chi_{\rm p}$ &    &    &  $\chi_{\rm w}$&   \\
 $\Delta$ &  $A$  & $\eta$   &  $\chi^2$ & $B$ &  $L_0$ &  $\chi^2 $ \\
\hline
 1.00 &  0.52(1) & 0.198(9)& 0.306 & 2.49(7) & 0.1(2) & 0.409 \\ 
 1.35 &  0.46(1) & 0.198(7)& 0.020 & 2.14(3) & 0.6(2) & 0.251 \\ 
 1.45 &  0.474(8) & 0.219(6)& 0.834 & 2.04(4) & 0.7(3) & 1.644 \\ 
 1.55 &  0.50(1) & 0.25(1) & 0.676 & 1.91(4) & 0.8(4) & 3.838 \\ 
 1.65 &  0.50(2) & 0.28(1) & 5.466 & 1.72(4) & 1.4(4) & 2.291 \\ 
\hline
\end{tabular}
\end{center}
\label{tb1}
\end{table}

In Fig.~\ref{fig5} we plot our results for $\chi_{\rm p}$ and $\chi_{\rm w}$ as a function of $L$. The fits are shown in Table \ref{tb1}.  The data appear to be consistent with a BKT transition around a critical disorder of $\sim\! 1.5$. Note in particular the excellent power law behavior of $\chi_{\rm p}$ for $\Delta \!\le\! 1.55$. The expected form for $\chi_{\rm w}$ is seen for $\Delta \le 1.35$, but it does not fit very well for $\Delta \approx 1.5$. From the rapid decay of $\chi_{\rm w}$ (confirmed by saturation in $\chi_{\rm p}$, not shown), the system is clearly no longer superconducting when $\Delta \!=\! 3$.

To obtain a good estimate of the critical disorder, $\Delc$, we assume that the forms (\ref{fitcp})-(\ref{fitcw}) hold close to the transition and that the deviation of the exponent $\eta$ and constant $B$ is linear in $\Delta$: 
\begin{eqnarray}
\eta & = & a (\Delta - \Delc) + 0.25 \label{eq:etalin} \\
B & = & b (\Delta - \Delc) + 2.0 \;. 
\label{eq:Blin} 
\end{eqnarray}
A joint fit of $\eta$ and $B$ to this form for $\Delta$ in the interval $[1.35,1.65]$ yields $\Delc \!=\! 1.53(4)$.

When $U \!=\! -\infty$ our model reduces to the $XY$ model in the pseudo-spin variables. As seen in Fig.~\ref{fig5}, $\Delta=1.65$ is still in the superconducting phase in the $XY$ limit.\cite{Ana04} Thus, the effect of fermions is to disorder the superconductor more quickly, as can be intuitively expected because of the increased entropy.

As discussed earlier, when $J_3 \!=\! 1$ and $\mu \!=\! 0$ the model has an additional $SU(2)$ pseudo-spin symmetry, as in the attractive Hubbard model.  Thus, due to the Mermin-Wagner theorem, superconductivity is only possible when $\mu \!\neq\! 0$. The $J_3 \!=\! 1$ model also has been studied earlier in the absence of disorder\cite{Osb02}, and a BKT transition was established using universal finite size scaling. We have extended these calculations to the disordered regime. We again fix the temperature at $T \!=\! 0.25$ and study the effects of disorder with $\mu \!=\! 1$.

\begin{figure}[t]
\vbox{
\includegraphics[width=7.7cm]{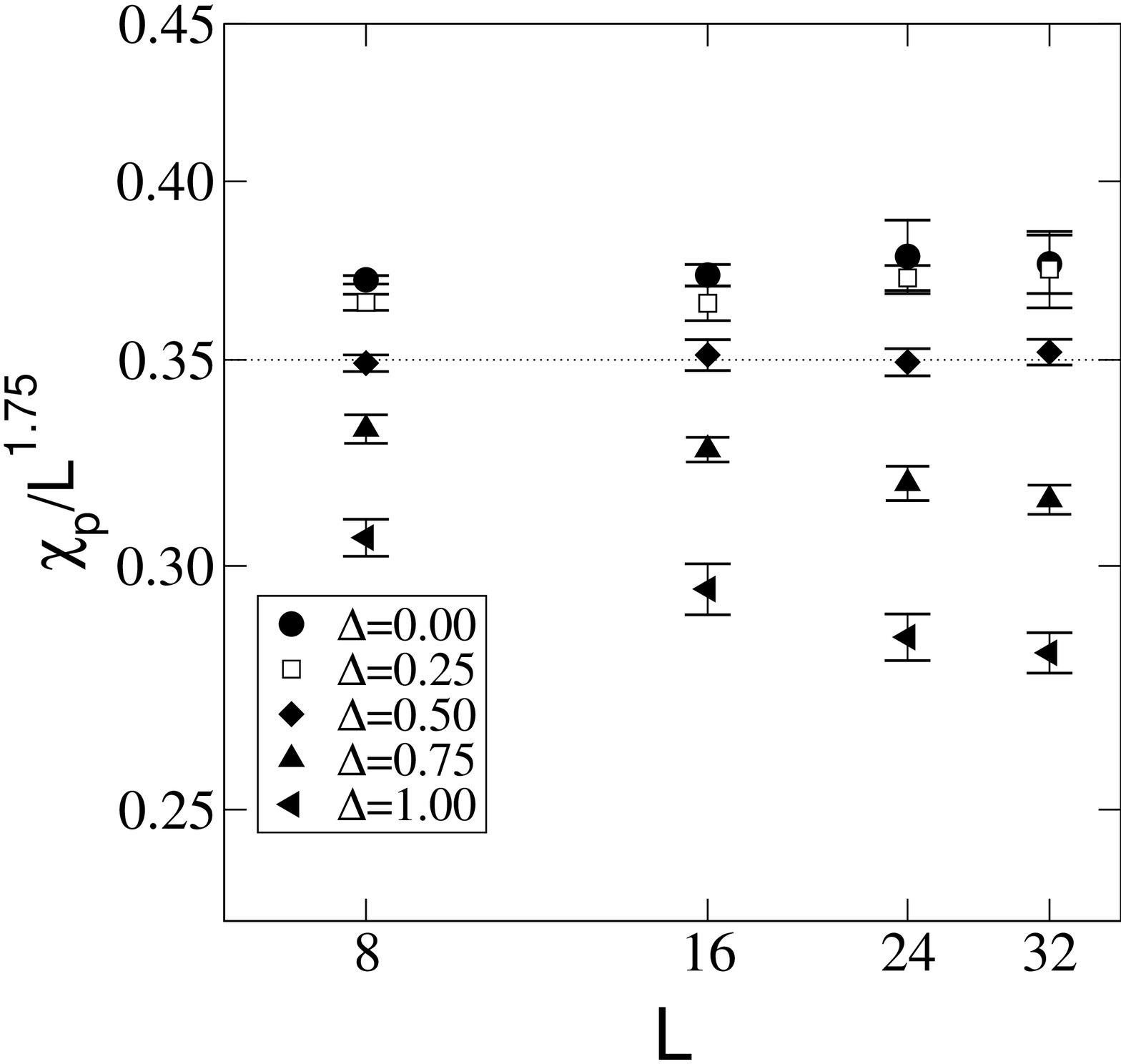}
\includegraphics[width=7.7cm]{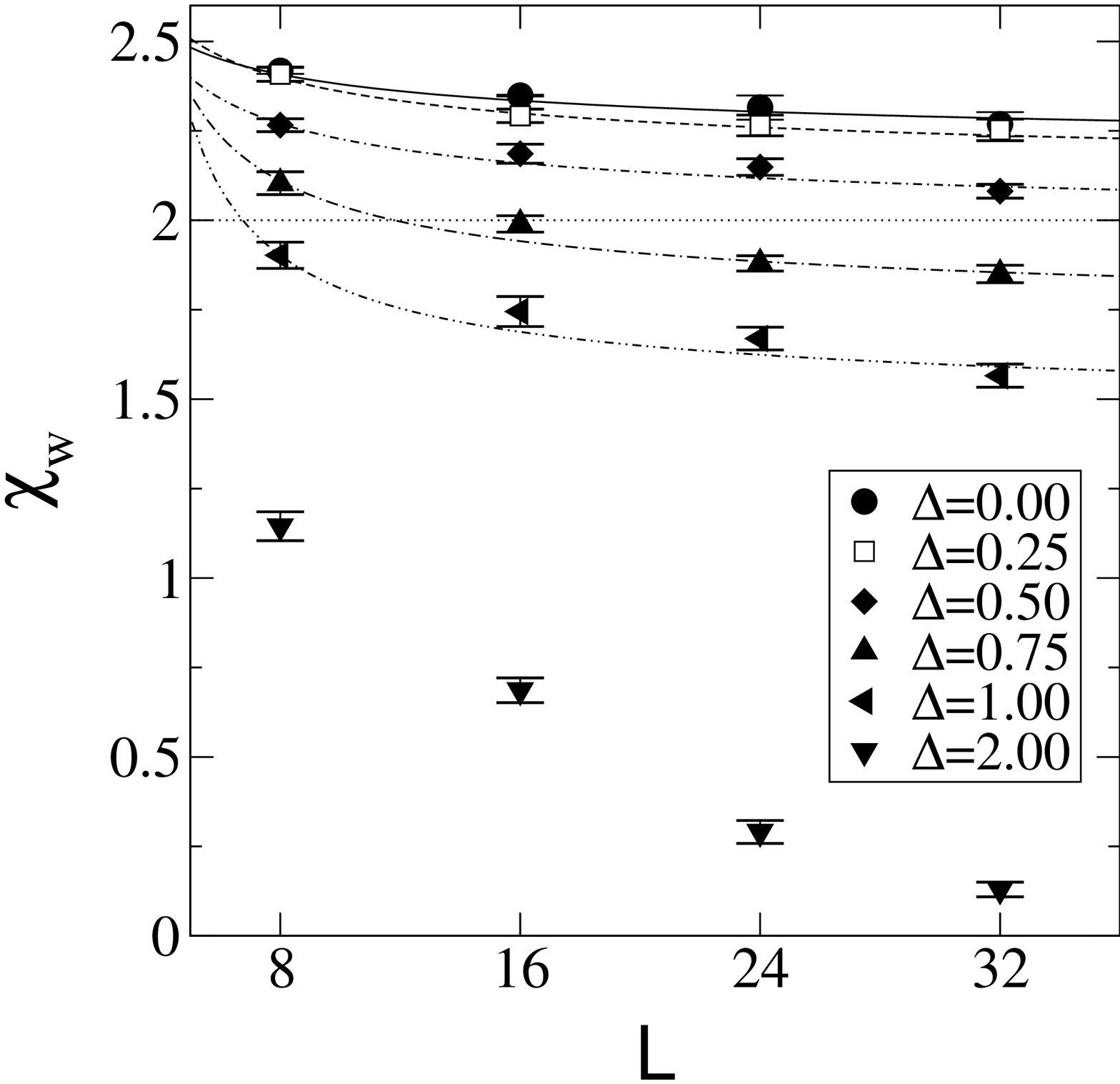}
}
\caption{Pair susceptibility divided by $L^{1.75}$ ($\chi_{\rm p}/L^{1.75}$, top, logarithmic scales) and winding number susceptibility ($\chi_{\rm w}$, bottom, linear scales) as a function of the lattice size for different magnitudes of disorder $\Delta$ at $J_3 \!=\! 1$, $U \!=\! -0.1$, $\mu \!=\! 1$, and $T \!=\! 0.25$.\label{fig6} The dashed horizontal lines are guides to the eye. Note the excellent BKT transition behavior. } \end{figure}

Fig.~\ref{fig6} gives our results and Table \ref{tb2} shows the corresponding fits.  The expected BKT transition behavior is indeed seen for $\Delta \!\le\! 0.5$.  Note that the values obtained for $\eta$ are constant, within statistical error, in this range. Thus to extract the critical disorder $\Delc$ we use only the $\chi_{\rm w}$ data. A fit of the $\Delta \!\le\! 0.5$ values to the form in  Eq.~(\ref{eq:Blin}) yields $\Delc \!=\! 0.13(5)$; the data for $\eta$ is consistent with this value. The critical disorder found here is much smaller than in the $J_3 \!=\! 0$ case above, indicating as expected that superconductivity is weaker when $J_3  \!=\!  1$.

\begin{table}[t]
\vspace{0.05cm}
\caption{Fitting results for the pair susceptibility ($\chi_{\rm p}$) and the winding number susceptibility ($\chi_{\rm w}$)
for $J_3 \!=\! 1$, $U \!=\! -0.1$, and $\mu \!=\! 1$ (Fig.~\ref{fig6}).  The fitting formulas are Eqs.~(\ref{fitcp})-(\ref{fitcw});
the $\chi^2$ given is per degree of freedom from the fit.}
\begin{center}
\begin{tabular}{l|| c c c| c c c  }
\hline
         &      & $\chi_{\rm p}$ &    &    &  $\chi_{\rm w}$&   \\
$\Delta$ &  $A$ & $\eta$  & $\chi^2$ & $B$ & $L_0$ &  $\chi^2$   \\
\hline
0.00 & 0.365(8) & 0.242(9) & 0.081  & 2.04(4) & 0.5(2) & 0.120 \\
0.25 & 0.354(8) & 0.235(9) & 0.383  & 1.97(3) & 0.8(2) & 0.277 \\
0.50 & 0.346(7) & 0.245(7) & 0.460  & 1.83(2) & 1.0(2) & 1.213 \\
0.75 & 0.36(1) & 0.29(1) & 0.446    & 1.58(2) & 1.8(3) & 1.952 \\
1.00 & 0.35(1) & 0.31(1) & 0.083    & 1.33(2) & 2.5(3) & 1.618 \\
\hline
\end{tabular}
\end{center}
\label{tb2}
\end{table}

\subsection{Fermionic degrees of freedom}

\begin{figure}[b]
\includegraphics[width=7.7cm]{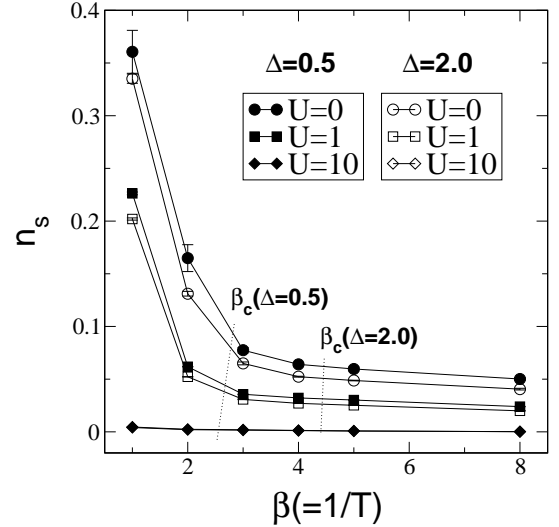}
\caption{Density of singly occupied sites versus inverse temperature for $\Delta \!=\! 0.5$ and $\Delta \!=\! 2$ at $J_3 \!=\! 0$, $\mu \!=\! 0$, and three different values of $U$. The dotted line indicates the approximate critical value of $\beta \!=\! 1/T$ where the system undergoes a BKT transition to a superconductor. The total particle density is $n \!=\! 1$ due to particle-hole symmetry.  \label{fig7}} 
\end{figure}

Since we are studying a strongly correlated electronic system with on-site attraction between spin-up and spin-down electrons, one might worry that the electrons have formed local pairs on each site and the model is essentially bosonic. Hence, it is important to look at observables that extract fermionic information in the model. One such observable is the density of singly occupied sites defined in Eq.~(\ref{ns}). In Fig.~\ref{fig7} we plot $n_s$ as a function of temperature when $J_3 \!=\! 0$ and $\mu \!=\! 0$. Since there is particle-hole symmetry at $\mu \!=\! 0$ we have $n \!=\! 1$. For this calculation we have fixed $\epsilon \!=\! \beta/M \!=\! 0.25$ for $\beta \!\geq\! 4$ and $M \!=\! 64$ for $\beta \!<\! 4$.  We show two different values of $\Delta$ -- one for weak disorder and the other for strong disorder -- at different values of $U$.  The data shown was obtained at $L \!=\! 16$; we have observed that the density does not vary much as the lattice size increases.

First we note that the density of singly occupied sites decreases significantly as the temperature is lowered for $U \!=\! 0$. When the disorder is small the pairs begin to break close to the critical temperature, while at large disorders the pairs break only at a temperature much higher than the critical temperature. Further, as $|U|$ is reduced from infinity, the background density of singly occupied sites increases as expected; the increase is significant for $T \!>\! 0.4$, while it is modest for smaller $T$. The background density approaches a constant as $T$ decreases, leading us to conclude that fermionic excitations do exist for all values of $T$. On the other hand the change in disorder $\Delta$ has almost negligible effect on $n_s$, especially near the phase transition (dotted lines in the figure). This leads us to conclude that the properties of the phase transition are most likely governed by a bosonic model such as the disordered quantum $XY$ model, especially in the strongly disordered case. Thus, our work gives credence to the ``dirty boson'' scenario.

\section{DIRECTIONS FOR THE FUTURE}

In this article we have studied the effects of disorder on a strongly correlated electronic system. Our model was known to be superconducting in the clean limit and our motivation in this work was to study the effects of disorder through a position dependent chemical potential.  Unfortunately, a naive extension of the meron cluster algorithm was found to be inefficient in the presence of a disordered chemical potential. Hence in this work we constructed a new and efficient algorithm by combining the the meron-cluster approach with the directed-loop algorithm. Earlier work on the directed-loop algorithm involved quantum spin systems and always was constructed in the spin representation. In this work we have shown that the algorithm can be constructed even in the cluster representation, which is essential in the fermionic system in order to solve the fermion sign problem. Our new algorithm was quite successful and allowed us to compute the pair and winding number susceptibilities accurately.

We found that disorder significantly suppresses superconductivity and the system undergoes a phase transition which appears consistent with the BKT universality class. Although this scenario has been expected,\cite{Wal94} our work is the first, as far as we know, to study the universal scaling predictions of the BKT transition in a fermionic system with disorder. We could go to lattices as large as $L \!=\! 32$ thanks to our new algorithm. We found that when $J_3 \!=\! 0$ superconductivity is stronger than when $J_3 \!=\! 1$.

We also found that there indeed are fermionic excitations in the system, but they are not affected by the disorder. The role of these fermions remains an interesting open question. For example do the background fermions form a Fermi-liquid in the weak disorder regime? If this is the case, then it would be interesting to ask whether the fermions become localized or do they remain extended. Is the phase transition between a superconductor and an insulator or whether it is a transition between a superconductor and a metal. Finally, although we have focused on attractive interactions in this work, we can study the repulsive model by setting $U$ positive.  In that case it is possible to add a chemical potential such that $\mu \leq U/2$ when $J_3 \!=\! 1$ without introducing a sign problem\cite{Shailesh}. These studies have the potential to increase the fermionic effects further.

\acknowledgments

We would like to thank A. Ghosal, N. Trivedi and R. Scalettar for helpful discussions. This work was supported in part by National Science Foundation grant DMR-0103003. The computations were performed on a Beowulf cluster funded by the same grant.

\bibliographystyle{apsrev}

\end{document}